\documentclass[aps,pra,twocolumn,superscriptaddress,showpacs]{revtex4-1}
\usepackage{graphicx}
\usepackage{amsmath}
\usepackage{epsfig}
\usepackage{bm}
\usepackage{multirow}

\newcommand{\beq}{\begin{eqnarray}}
\newcommand{\eeq}{\end{eqnarray}}

\begin{document}

\title{Cracking the invisible cloak by using the temporal steering inequality%
}
\author{Shin-Liang Chen}
\affiliation{Department of Physics and National Center for Theoretical Sciences, National
Cheng Kung University, Tainan 701, Taiwan}
\author{Ching-Shiang Chao}
\affiliation{Department of Physics and National Center for Theoretical Sciences, National
Cheng Kung University, Tainan 701, Taiwan}
\author{Yueh-Nan Chen}
\affiliation{Department of Physics and National Center for Theoretical Sciences, National
Cheng Kung University, Tainan 701, Taiwan}
\email{yuehnan@mail.ncku.edu.tw}
\date{\today }

\begin{abstract}
Invisible cloaks provide a way to hide an object under the detection of
waves. A perfect cloak guides the incident waves through the cloaking shell
without any distortion. In most cases, some important quantum degrees of
freedom, e.g. electron spin or photon polarization, are not taken into
account when designing a cloak. Here, we propose to use the temporal
steering inequality of these degrees of freedom to detect the existence of
an invisible cloak.
\end{abstract}

\maketitle

\section{Introduction}

Invisible cloaks based on the transformation design method (TDM) has
attracted great attentions in the past decade~\cite%
{Pendry06,Leonhardt06,Schurig06,Milton06,Alu08,Farhat08,Zhang11,Zhang08,Guenneau12}%
. The main idea of the TDM is to perform the coordinate transformation on
the wave equation of the corresponding cloaking wave to create the hiding
region. To keep the form of the equation invariant, the metric tensors are
combined with the specific parameters, which are usually the properties of
the material of the cloaking shell. For instance, the TDM for
electromagnetic waves~\cite{Pendry06,Leonhardt06,Schurig06} reinterprets the
effect of the coordinate transformation as conductivity and permeability in
the original non-transformed system. Similarly, cloaking of matter waves~%
\cite{Zhang08,Lin09,Chen12,Chang14} requires a proper design of the
effective mass and potential of the cloaking shell. There are also other
kinds of cloak, such as cloaking of elastic waves~\cite%
{Milton06,Cummer07,Norris08,Brun09,Stenger12}, liquid waves~\cite%
{Farhat08,Zhang11}, heat flows~\cite{Guenneau12,Schittny13}, etc. Waves
incident onto the cloak designed by the TDM are perfectly guided through the
cloaking shell without any scattering and distortion.

Einstein-Podolsky-Rosen (EPR) steering~\cite%
{Schrodinger35,Reid89,Wiseman07,Cavalcanti09} is one of the quantum
correlations that allows one party to remotely prepare some specific states for
the other party via choosing different measurement settings. The degree of
the non-locality of EPR-steering is stronger than the entanglement but
weaker than the Bell non-locality~\cite{Wiseman07}. EPR steering can be
verified via the steering inequalities~\cite{Cavalcanti09}, which are built
on the fact that the correlations cannot be explained by the local hidden
state model. Apart from the correlations between two (or more) parties,
quantum correlations may also occur in single party at different times. For
example, Leggett and Garg derived an inequality~\cite{LG85,Emary14} under
the assumption of macroscopic realism and non-invasive measurement. It can
be used to verify the quantum coherence of a macroscopic system under the
weak measurements~\cite{Palacios-Laloy10}. Recently, a temporal analog of
the steering inequality --- the temporal steering inequality~\cite{Chen14}
--- also focuses on the correlations of a single party at different times.
Moreover, the classical bound of temporal steering inequality is found to
have deep connection with the quantum cryptography.

In this work, we consider a two-dimensional cylindrical cloak designed by
the TDM, and the waves can be perfectly cloaked in the entire space. For
concreteness, we consider the invisible cloak of the electromagnetic waves
and the electron matter waves. Since the spin of the incident matter waves
(e.g., an electron) may interact with the hiding object when passing through
the cloaking shell, we assume the incident particle experiences a coherent
coupling. Secondly, we assume the polarizations of the incident
electromagnetic waves suffer a phase damping when passing through the
cloaking shell. The feature of temporal steering inequality~is that the
temporal steering parameter always maintains the maximal value if the wave
does not interact with other ancillary systems (or environment). Our results
show that the temporal steering parameter of the incident waves varies with
the traveling time in the cloaking shell, and therefore the invisible cloak
can be cracked by using the temporal steering inequality.

\section{TRANSFORMATION DESIGN METHOD FOR WAVES}


\begin{figure}[th]
\emph{\includegraphics[width=8.5cm]{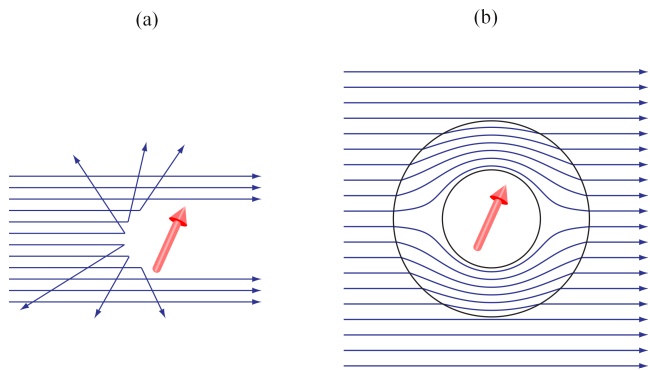} }
\caption{(Color online) (a) An object is observed by the detection of the
scattering waves. (b) A cloak designed by the TDM perfectly guides the
incident waves passing through the cloaking shell.}
\label{cloak}
\end{figure}

One of the crucial points in the TDM for waves is to perform the appropriate
coordinate transformation on the spatial (time-independent) wave equation
from the coordinate system $q$ to $q^{\prime }$, and keep the form invariant
\begin{equation}
\begin{aligned} &\nabla^2\Psi(q)+k^2\Psi(q)=0\\ \mapsto~ &\nabla^{\prime
2}\Psi(q^\prime)+k^{\prime 2}\Psi(q^\prime)=0, \end{aligned}
\end{equation}%
where $k^{\prime }$ reinterprets the effect of the coordinate transformation
in the properties of the material in the cloaking shell.

\begin{figure}[th]
\emph{\includegraphics[width=8.5cm]{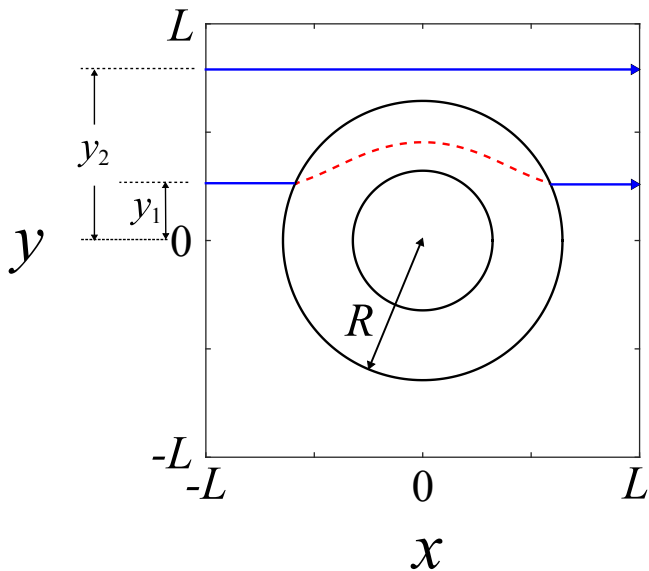} }
\caption{(Color online) The trajectory of the incident particle (photons or
electrons) from the current density in the classical limit. The direction of
the incident plane waves $\Psi =e^{i(kx-\protect\omega t)}$ is along $+x$.
Here, $x$ and $y$ are in units of $(1/k)$}
\label{path}
\end{figure}

The behavior of the incident waves can be visualized through the current
density $\mathbf{J}$, with the continuity equation
\begin{equation}
\nabla \cdot \mathbf{J}=-\frac{\partial \sigma }{\partial t},
\end{equation}%
where $\sigma =\Psi ^{\ast }\Psi $ is the probability density of the wave
function. The incident plane wave $\Psi =e^{i(kx-\omega t)}$ could be the
electromagnetic wave (photons) or the matter wave (electrons). One can use
the relation, $\mathbf{J}=\sigma \mathbf{v}$, to obtain the classical
trajectory of the incident particle~\cite{Sakurai93}. In classical limit,
the velocity vector $\mathbf{v}$ is tangent to the particle trajectory.
Therefore, the trajectory of the incident waves can be obtained from the
current density (Fig.~\ref{path}).

Moreover, it is necessary to estimate the time interval $t_\text{s}$ of the
incident particle staying inside the cloaking shell. The phase of the
incident wave after passing through the shell of a perfect cloak should be
the same as that traveling in free space. Thus, the time intervals for
different trajectories should be the same. As an example, we consider two
trajectories, representing the path that the particles travel from $x=-L$ to
$x=+L$ with and without passing through the shell, respectively (Fig.~\ref%
{path}). The time interval $t_\text{s}$ can then be easily obtained
\begin{equation}
t_\text{s}=\frac{2L}{v}-\frac{2L-2\sqrt{R^{2}-y_{1}^{2}}}{v},
\label{ts}
\end{equation}
where $v=\omega/k$ is the velocity of the incident particle outside the cloak.

\section{TEMPORAL STEERING INEQUALITY}

In this section, we briefly describe the concept of the temporal steering
inequality~\cite{Chen14}. Consider a two-level system sent into one of the
channels $\lambda $ with the probability $q_{\lambda }$. During the
transmission, there are two observers, Alice and Bob. Firstly, Alice
performs the measurement on the system at time $t_{\text{A}}$ along the
basis $i$ with the outcomes $A_{i,t_{\text{A}}}=a$. Then, the system is
suffered from the influence of the channel for a time interval before Bob
receives it. When Bob receives the system at time $t_{\text{B}}$, he obtains
the outcomes $B_{i,t_{\text{B}}}=b$ by performing the measurement along the
same setting $i$. If Alice's choice of measurement has no influence on the
state that Bob receives, the following temporal steering inequality holds
\begin{equation}
S_{N}^{\text{T}}\equiv \sum_{i=1}^{N}E\left[ \langle B_{i,t_{\text{B}%
}}\rangle _{A_{i,t_{\text{A}}}}^{2}\right] \leq 1,  \label{TSI}
\end{equation}%
and the bound that quantum mechanics gives is
\begin{equation}
S_{N}^{\text{T}} \leq N
\end{equation}
where $N$(= 2 or 3) is the number of the mutually unbiased measurements that
Bob implements on the system, and
\begin{equation}
E\left[ \langle B_{i,t_{\text{B}}}\rangle _{A_{i,t_{\text{A}}}}^{2}\right]
\equiv \sum_{a=\pm 1}P(A_{i}=a)\langle B_{i,t_{\text{B}}}\rangle _{A_{i,t_{%
\text{A}}}}^{2},
\end{equation}%
with
\begin{equation}
P(A_{i}=a)\equiv \sum_{\lambda }q_{\lambda }P_{\lambda }(A_{i}=a).
\end{equation}%
Here, Bob's expectation value conditioned on Alice's result is defined as
\begin{equation}
\langle B_{i,t_{\text{B}}}\rangle _{A_{i,t_{\text{A}}}}\equiv \sum_{b=\pm
1}bP(B_{i,t_{\text{B}}}=b|A_{i,t_{\text{A}}}=a).
\end{equation}%
Here, we would like to use two measurement settings, the $\hat{X}$ and $\hat{%
Z}$ bases, rather than three. Since three measurement settings are
sufficient to perform the quantum state tomography, using the temporal
steering inequality thus requires fewer resources. We use one of the
features of the temporal steering parameter $S_{N}^{\text{T}}$ in Eq.~(\ref%
{TSI}) to detect the quantum cloak: If the system does not suffer any
interaction, quantum mechanics predicts that $S_{2}^{\text{T}}$ always
maintains the maximal value $2$. If $S_{2}^{\text{T}}$ varies with time, the
system is subject to some dynamics.

\section{Cracking electromagnetic cloak by using the temporal steering
inequality}

\begin{figure}[th]
\emph{\includegraphics[width=8cm]{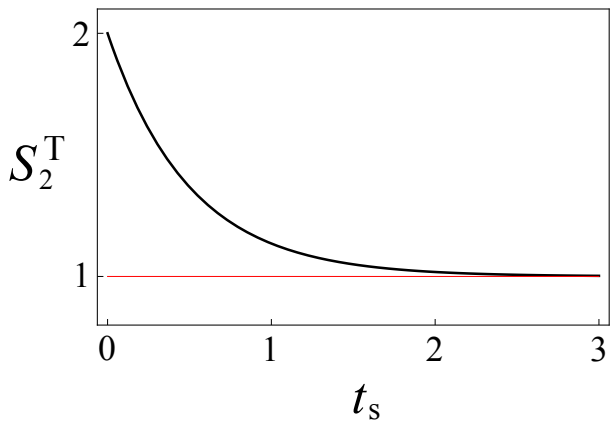} }
\caption{(Color online) The dynamics of the temporal steering parameter $%
S_{2}^{\text{T}}$ of the polarizations of the incident photons when
suffering a phase damping inside the cloaking shell. The horizontal red line
represents the classical bound of the temporal steering inequality. In
plotting the figure, the time $t_\text{s}$ is in units of $1/\protect\gamma$. }
\label{TS2_pure_dephasing}
\end{figure}

We assume the incident photons suffer a phase damping with decay rate $%
\gamma $ when traveling through the cloaking shell. The state of the
polarizations can be described by the density matrix $\rho =\sum_{i,j}\rho
_{ij}|i\rangle \langle j|$, where $|i\rangle $, $|j\rangle $ $\in $ $%
\{|H\rangle ,|V\rangle \}$ with $|H\rangle $ and $|V\rangle $ being the
horizontal and vertical polarizations, respectively. The initial state is
prepared in the maximally mixed state
\begin{equation}
\rho (t=0)=\frac{1}{2}(|H\rangle \langle H|+|V\rangle \langle V|)=%
\begin{pmatrix}
1/2 & 0 \\
0 & 1/2%
\end{pmatrix}%
.  \label{IC}
\end{equation}%
The evolution of the polarizations inside the cloaking shell can be obtained
by solving the following Markovian master equation with Lindblad form~\cite%
{Lindblad76,Gorini76}
\begin{equation}
\frac{\partial \rho _{1}(t)}{\partial t}=\frac{\gamma }{4}[2\sigma _{z}\rho
_{1}(t)\sigma _{z}-\sigma _{z}^{2}\rho _{1}(t)-\rho _{1}(t)\sigma _{z}^{2}],
\label{MME}
\end{equation}%
where $\sigma _{z}$ is the Pauli-$z$ matrix. From Eqs.~(\ref{TSI}), (\ref{IC}%
), and (\ref{MME}) the steering parameter can be obtained straightforwardly%
\begin{equation}
S_{2}^{\text{T}}=1+e^{-2\gamma t_\text{s}},
\end{equation}%
where $t_\text{s}$ is defined in Sec.II.
Here, the two bases are $\{|H\rangle ,|V\rangle \}$ and $\{(|H\rangle
+|V\rangle )/\sqrt{2},~(|H\rangle -|V\rangle )/\sqrt{2}\}$. The dynamics of
the temporal steering parameter $S_{2}^{\text{T}}$ of the polarizations is
plotted in Fig.~\ref{TS2_pure_dephasing}. We can see that the temporal
steering parameter $S_{2}^{\text{T}}$ varies with time inside the shell ($%
t_\text{s}$). Therefore, the electromagnetic cloak is cracked by using the
temporal steering inequality.

\section{Cracking quantum cloak by using the temporal steering inequality}

\begin{figure}[th]
\emph{\includegraphics[width=8cm]{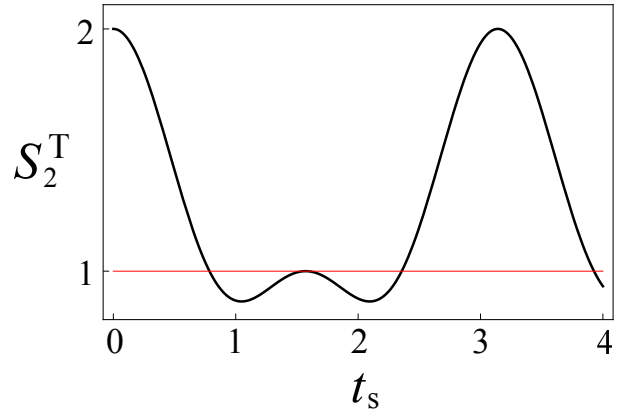} }
\caption{(Color online) The dynamics of the temporal steering parameter $S_2^%
\text{T}$ of the spin of the incident electron when it passes through the
cloaking shell. The horizontal red line represents the classical bound of
the temporal steering inequality. In plotting the figure, the time $t_\text{s}$ is
in units of coupling strength $J$, and $\hbar=1$.}
\label{TS2_coupling}
\end{figure}

Although the cloaking of two and three dimensional spinor have been proposed~%
\cite{Lin10,Lin11}, the material of such a cloak is extremely hard to be
realized. In this section, we use the temporal steering parameter to detect
the dynamics of the spin of a quantum particle inside the cloaking shell.
For simplicity, we consider the incident matter wave with the spin-1/2
degree of freedom, e.g. electrons. We further assume the spin of the
electron experiences the coherent coupling from the ancillary spin hidden
inside the cloaking shell. The state of the incident spin can be described
as $\rho =\sum_{i,j}\rho _{ij}|i\rangle \langle j|$, where $|i\rangle $, $%
|j\rangle $ $\in $ $\{|\uparrow \rangle ,|\downarrow \rangle \}$ with $%
|\uparrow \rangle $ and $|\downarrow \rangle $ being the spin-up and
spin-down state, respectively. The interaction Hamiltonian can be written as
$H=\hbar J(\sigma _{+}^{1}\sigma _{-}^{2}+\sigma _{+}^{2}\sigma _{-}^{1})$,
where $\sigma _{+}^{i}$ and $\sigma _{-}^{i}$ are the raising and lowering
operators of the $i$ th spin, and $\hbar J$ is the coupling strength. The
evolution of the entire system inside the cloaking shell can be obtained by
the quantum Liouville equation
\begin{equation}
\frac{\partial \rho _{12}(t)}{\partial t}=\frac{1}{i\hbar }\left[ H,\rho (t)%
\right] .  \label{QLE}
\end{equation}%
The state of the incident electron $\rho _{1}(t)$ can be obtained by tracing
out the ancillary electron, i.e. $\rho _{1}(t)=\text{Tr}_{2}\left[ \rho
_{12}(t)\right] $. We choose the initial state as
\begin{equation}
\rho _{12}(t=0)=%
\begin{pmatrix}
1/2 & 0 \\
0 & 1/2%
\end{pmatrix}%
\otimes
\begin{pmatrix}
0 & 0 \\
0 & 1%
\end{pmatrix}%
.
\end{equation}%
The temporal steering parameter $S_{2}^{\text{T}}$ is then written as
\begin{equation}
S_{2}^{\text{T}}=\frac{1}{4}\left[ 5+2\cos (2Jt_{s})+\cos (4Jt_{s})\right] ,
\end{equation}%
where the two bases are Pauli $\hat{X}$ and $\hat{Z}$. From Fig.~\ref%
{TS2_coupling}, we can see that $S_{2}^{\text{T}}$ varies with time,
indicating the incident electron is not traveling through the free space.
Therefore, the quantum cloak is cracked by using the temporal steering
inequality.

\section{CONCLUSION}

One may notice that there are other ways to crack the invisible cloak. A
simple way is to detect whether the direction of the spin (or
depolarization) is changed. However, this method requires the measurement
direction of the receiver to be synchronized with that of the sender. In the
temporal steering scenario, there is no such constraint, i.e. the steering
inequality still holds even if the bases are not synchronized~\cite{Chen14}.
Another way to crack the cloak is the quantum state tomography. In this
case, one has to use three bases (for qubit system) to perform the
tomography, whereas one only needs two bases for the temporal steering
inequality. Besides, one may also use the degree of entanglement to detect
the cloak: preparing initially the entangled pair, sending one of them into
the shell, and measuring the degradation of the entanglement. However, this
would require more quantum resources and cannot detect the coherent-coupling
case in Fig. 4. In conclusion, the temporal steering inequality provides a
relative simple and efficient way to crack the invisible cloak.

\begin{acknowledgements}
The authors would like to thank D.-H. Lin for the useful discussion. This work is supported partially by the National Center for Theoretical Sciences and Ministry of Science and Technology, Taiwan, grant number MOST 103-2112-M-006-017-MY4.
\end{acknowledgements}

%


\end{document}